\documentclass[11pt]{article}
\usepackage{multicol}
\usepackage{graphicx}
\usepackage{amssymb,bm,mathrsfs,bbm,amscd}
\usepackage[tbtags]{amsmath}
\usepackage{lastpage}
\usepackage{CJK}
\usepackage{indentfirst}
\date{}

\title{ 1D Longitudinal Beam Dynamics of Laser Plasma Wakefield Accelerator }
\author{ Xiongwei Zhu \\
{Institute of High Energy Physics, Chinese Academy of Sciences,}\\
 {P.O.Box 918, Beijing 100049, China}}

\begin{document}

\maketitle

\begin{abstract}

 In this paper, we get the 1D approximate analytical solution of the plasma electrostatic wake driven by the laser, and get the oscillating frequency of the wake modified by the nonlinear high order terms, and find that the frequency depends on the oscillating amplitudes of the wake and the laser which is the general nonlinear phenomenon. Finally we analyze the longitudinal beam dynamics in this electrostatic wake, and find that the high order terms don't change the topology of the longitudinal phase space.

\end{abstract}



\section{Introduction}

  Laser plasma wakefield accelerators (LWFA), in which the plasma wakefield is excited by an intense laser to accelerate the particles, have demonstrated accelerating gradient of hundreds of GV/m, and have become the highlights of advanced accelerator concept\cite{1}. In this paper, we get the 1D analytical solution of the plasma electrostatic wake driven by the laser, and get the modified oscillating frequency of this wake. Finally we analyze the longitudinal beam dynamics in this electrostatic wake, and find that the high order terms don't change the topology of the longitudinal phase space.

\section{Basic equations for LWFA}

   We consider the homogeneous unmagnetized plasma. Laser travels in the $z$ direction with the initial frequency $\omega_{0} \gg \omega_{p}$ , $\omega_{p}$ is the electron plasma frequency, and the group and the phase speed  $v_{g} \approx v_{p} \approx c$
   ($c$  is the light velocity ) . We adopt one dimensional theory, the LWFA system is governed by the following set of equations\cite{2}

\begin{equation}
\frac{\partial n_{e}}{\partial t} + \frac{\partial n_{e} v_{\parallel}}{\partial z} = 0,
\end{equation}

\begin{equation}
\hat{z} \times \frac{\partial \vec{E}}{\partial z} = - \frac{1}{c} \frac{\partial \vec{B}}{\partial t},
\end{equation}

\begin{equation}
\hat{z} \times \frac{\partial \vec{B}}{\partial z} = - \frac{4 \pi e}{c} n_{e} \vec{v} + \frac{1}{c} \frac{\partial \vec{E}}{\partial t},
\end{equation}

\begin{equation}
\frac{\partial E_{\parallel}}{\partial z} = - 4 \pi e (n_{e} - n_{0}),
\end{equation}
where,   $\vec{E},\vec{B}$ are the electromagnetic fields,   and $\vec{v}$ is the electron velocity. We introduce the coordinates $ \xi =
k_{p} ( z - ct ) ( k_{p} = \omega_{p} / c )$, and normalized scalar and vector potentials $\phi(\xi)$ and $ a(\xi) $ . Such that
$E_{\parallel} = - \frac{m c^2}{e}\frac{\partial \phi}{\partial \xi}, E_{\perp} = \frac{m c^2}{e} \frac{\partial a}{\partial\xi}$  , where   $\perp$ and $\parallel$ refer to the components perpendicular and parallel to $\hat{z}$ . From the equations above, we can get the equations about $ \phi (\xi) $ and $ a(\xi) $ ,

\begin{equation}
\frac{\partial^{2} a}{\partial^{2} \xi} = \frac{n_{e}}{n_{0}} \frac{ a }{\gamma},
\end{equation}

\begin{equation}
\frac{\partial^{2} \phi}{\partial^{2} \xi} = \frac{n_{e}}{n_{0}} - 1,
\end{equation}
where $k_{p} = \frac{2 \pi}{\lambda_{p}}$ is the electron plasma wave number, $\gamma$ is the relativistic factor of electron. From the Hamiltonian of the electron and the continuity equation, one can get the following relation

\begin{equation}
\frac{n_{e}}{n_{0}} = \frac{\gamma}{1 + \phi} = \frac{1 + a^2 + (1+\phi)^2}{2 (1+\phi)^2}.
\end{equation}

Finally, we get the 1D self-consistent equations for LWFA system

\begin{equation}
\frac{d^{2} a}{d^{2} \xi} = \frac{a}{1 + \phi},
\end{equation}

\begin{equation}
\frac{d^2 \phi}{d^2 \xi} = \frac{1}{2} (\frac{1 + a^2}{(1+\phi)^2} -1 ),
\end{equation}
the difference between the equations above and that in \cite{2} is that we use the direct differential and the equations are self-consistent, we don't need other variables to solve the equations.

\section{The oscillating frequency of the wake}

In this paper, we assume that $a$ is unchanged and $\phi$ is in the vicinity of zero (small variable) $\phi \propto \epsilon $ to get the analytical solution of the electrostatic plasma wake. We expand the equation (9) into the form

\begin{equation}
\phi'' +  \phi - \frac{3}{2} \phi^2 + 2 \phi^3 - \frac{5}{2} \phi^4 + ...
= \epsilon^2 a^2 ( 1 - 2\phi + 3\phi^2 - 4\phi^3 + 5\phi^4 + ...).
\end{equation}
Using KBM method\cite{3}, one expand as following

\begin{equation}
\phi = u cos\theta + \sum_{1}^{N} \epsilon^n \phi_{n}(u,\theta) + O(\epsilon^{n}).
\end{equation}
The equations of $u$ and $\theta$ are

\begin{equation}
\frac{d u}{d t} = \sum_{1}^{N} \epsilon^n A_{n} (u),
\end{equation}

\begin{equation}
\frac{d \theta}{d t} = \sum_{1}^{N} \epsilon^n \epsilon^n \theta_{n}(u).
\end{equation}
From the equation (10) , we remove the secular terms to obtain the first order solution
\begin{equation}
A_{1} = 0, \theta_{1} = 0
\end{equation}

\begin{equation}
\phi_{1} = \frac{3 u^2}{4} - \frac{u^2}{4} cos2\theta,
\end{equation}
and also the second order solution

\begin{equation}
A_{2} = 0, \theta_{2} = a^2 - \frac{9 u^2}{32},
\end{equation}

\begin{equation}
\phi_{2} = a^2 + \frac{11 u^3}{128} cos3\theta.
\end{equation}
So the solution of the electrostatic plasma wake potential to second order is

\begin{equation}
\phi = u cos\theta + \epsilon ( \frac{3 u^2}{4} - \frac{u^2}{4}
cos2\theta ) + \epsilon^2 ( a^2 + \frac{11 u^3}{128} cos3\theta),
\end{equation}
where $\theta = k_{p} (1 + a^2 - \frac{9 u^2}{32})(z - ct)$. In (18), if we reserve the zero order term only, then  , the wake oscillating frequency is the usual electron plasma frequency $ \omega_{p} $. But the wake oscillating frequency is modified as (19), if we include the high order terms, the modified frequency $ \omega_{p}' $ is related to the oscillating amplitude

\begin{equation}
\omega_{p}' = \omega_{p} ( 1 + a^2 - \frac{9 u^2}{32} ).
\end{equation}
The modified frequency depends on the square of the amplitude, this is the general nonlinear phenomenon\cite{4} and can be extended to the more common case in plasma physics.

\section{The longitudinal beam dynamics}

Now£¬ we begin to analyze the longitudinal beam dynamics in the electrostatic potential  . In doing so, the expression of   should be expressed as $ \theta = k_{p} (1 + a^2 - \frac{9 u^2}{32})(z - v_{p} t), v_{p} < c $. We incorporate the new variable $ s = k_{p} z $,
the longitudinal equations of motion are

\begin{equation}
\frac{d \gamma}{d s} = \frac{\partial \phi}{\partial \theta},
\end{equation}

\begin{equation}
\frac{d \theta}{d s} = 1 - \frac{\beta_{p}}{\beta},
\end{equation}
where $\beta_{p} = v_{p} / c, \beta = ( 1 - \gamma^{-2})^{\frac{1}{2}}$. The function $\phi$ is shown in the figure 1,

\begin{center}
\includegraphics[width=10 cm]{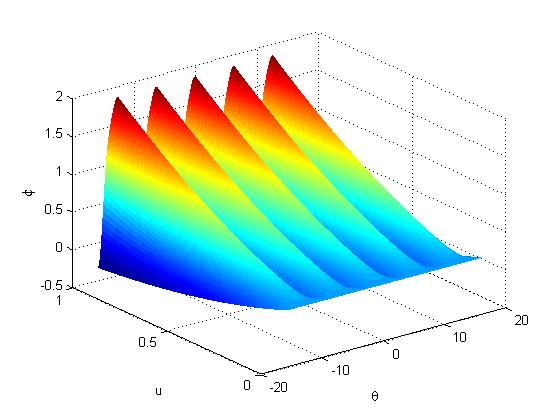}

{Figure 1. The function $\phi$ ( $a = 0.5$ ).}
\end{center}

The fixed points of the longitudinal equations of motion are $ \gamma = \gamma_{p}, \theta = n \pi $ ( $n$ is integer ). From the Jacobian matrix of (20) and (21), 
the characteristic value $\lambda$  satisfies $\lambda^2 = \frac{1}{\gamma_{p} \beta_{p}^2} \frac{\partial^2 \phi}{\partial^2 \theta}$. It is obvious that $\theta = 0$ is the stable point and $\theta = -\pi, \pi$ are the saddle points. If we remove the second and third terms in (18), we will get the same conclusion. So the high order terms in the electrostatic potential don't change the topology of the longitudinal phase space. Figure 2 gives the longitudinal phase space near the zero phase, when $u = 0.5, \beta_{p} = 0.98$.

\begin{center}
\includegraphics[width=10 cm]{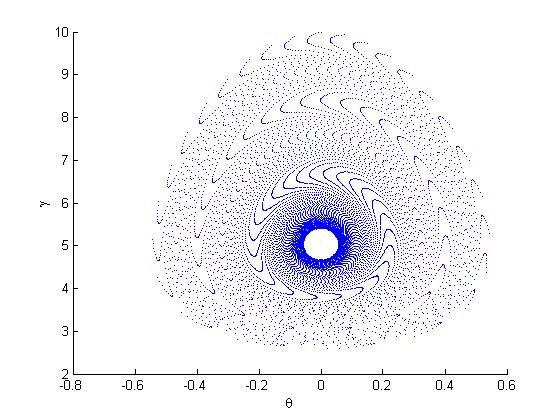}

{Figure 2.The longitudinal phase space near zero.}
\end{center}

\begin{center}
\includegraphics[width=10 cm]{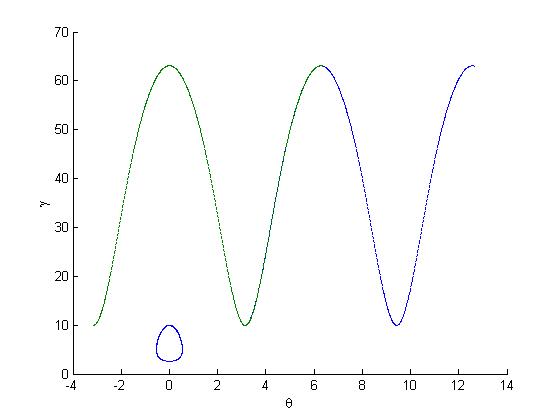}

{Figure 3. The whole longitudinal phase space.}
\end{center}

\end{document}